\documentclass[12pt,english]{iopart}
\usepackage[english]{babel}
\usepackage{graphicx}
\usepackage{moreverb}
\usepackage{hyperref}
\usepackage{color}
\usepackage[normalem]{ulem}
\newcommand{\unitto}[1]{\raisebox{.5ex}{{\small #1}}}
\newcommand{\exampleFile}[1]{
\texttt{#1}
}

\begin{document}

\title[\texttt{pygiftgenerator}: Moodle-based quizzes in
  python]{\texttt{pygiftgenerator}: A \texttt{python} module designed
  to prepare Moodle-based quizzes}

\author{Jon S\'aenz$^1$, Idoia G Gurtubay$^2$, Zunbeltz Izaola$^3$ and Gabriel
  A L\'opez$^1$ }

\address{$^1$ Department of Applied Physics II, University of 
the Basque Country (UPV/EHU), Sarriena Auzoa z/g,
  48940-Leioa, Spain.\\ $^2$ Department of Condensed Matter Physics,
  University of the Basque Country (UPV/EHU), Sarriena Auzoa z/g, 48940-Leioa, Spain.\\
  $^3$Escuela Universitaria de Ingenier\'\i a Dual,
  Instituto de M\'aquina Herramienta (IMH) Azkue auzoa 1, Elgoibar, Spain.}
\ead{jon.saenz@ehu.eus}
\begin{abstract}

We present \texttt{pygiftgenerator}, a \texttt{python} module for systematically preparing a large number
of numerical and multiple-choice questions for Moodle-based 
quizzes oriented to students' formative evaluation.
The use of the module is illustrated
by means of examples provided with the code and they cover topics from Mechanics, Electromagnetism, Thermodynamics and Modern Physics.
 The fact that \texttt{pygiftgenerator} relies on a well-established computer language which allows to combine and reuse functions in order to solve complex problems, makes it a very robust tool. Simply by changing the input parameters,
a large question bank 
with solutions even to complex physical problems can be generated.
Thus, it is a powerful alternative   
to the calculated  and multiple choice questions which can be written directly in the Moodle platform.
The module writes the questions to be imported into Moodle in a simple and human-readable ASCII output using the GIFT format,
which enables
html definitions for URLs for importing figures, or for simple text  formatting (sub/superindices or Greek letters) for equations and units. This format also admits \LaTeX\ typing for complex equations. 

\end{abstract}


\submitto{\EJP}

\maketitle

\section{Introduction}
\label{sec:intro}

The introduction of the European Credit Transfer System (ECTS)
criteria \cite{ECTS-2017}, which are the standards accepted by all
universities included in the European Higher Education Area (EHEA) to
guarantee the convergence of the different education systems in
Europe, and the launching of the so-called Bologna Process
\cite{BP-2015} has introduced a revolution in the teaching-learning
process in European universities.  The teaching-learning process has switched
from a teaching-centered education to a learning-centered one,
requiring that i) lecturers change traditional teaching methodologies
to give students the leading position and ii) students become active
learners, taking a larger commitment in their learning process, a role
which enables them reach the competences and learning outcomes
designed in the new curricula.  To this end, all the universities have
adopted new technologies specially suited to offering new online
teaching and learning environments to both
students and lecturers.  There is no doubt that this change of
paradigm would not have succeeded without the introduction of all
these online resources.

Recent reviews show that online learning is optimal when it is combined with the face-to-face modality
and this effect is statistically
significant~\cite{usreport}. Nevertheless, lecturers should not be
tempted to use online resources just as mere vehicles that deliver
instruction \cite{Kozma-1994,Clark-1994} and should design them to
make the most out of their capabilities. Doing so has demonstrated
that the use of active learning techniques contributes to improving
the outcomes of the learning process~\cite{Borondo-2014, Freeman-2014,
  ICERI-2017,IATED-2018,EDULEARN-2018,EuroSTL-2019}.  Online platforms
such as Moodle \cite{moodle}, allow students
to tackle different activities: downloading lecture
notes at home, exchanging ideas in discussion forums, uploading
assignments, peer-instruction activities, chat-rooms, surveys,
quizzes, plugins to virtual online classrooms and many others.
In the last decade there has been a significant proliferation of online courses in many universities~\cite{broadbent_2018, liagkou_2018}, although there are not so many studies devoted to massive assessment tools~\cite{corrigan_et_al_2015}.


The use of Moodle-based quizzes for formative evaluation has been
considered for years. In a recent study~\cite{sithara_et_al_2019}, the
authors focused in the optimization of Moodle quizzes for
online asessments. They showed that there are many important reasons
to consider Moodle-based tests as an instrument during the learning process
at the university level. The first reason is that they are effective
learning tools from the point of view of the knowledge acquired by the
students during the course. An important aspect in this regard is the
automated feedback they offer, which is able to engage students better
than delayed feedback. On the other hand, automated grading by Moodle
according to a well-constructed question bank saves significant
amounts of time that the instructors should otherwise invest in
manually correcting exams and calculating the marks obtained by
students. Automatic grading allows that this scarce resource (time)
can be devoted to other teaching activities such as preparation of
experiments, classroom demonstrations, new handouts, and so on. On the
other hand, the ability to randomize questions and answers inside
them allows to produce tests which are relatively robust against
plagiarism~\cite{sithara_et_al_2019}. This is an important
characteristic in case some of the quizzes are unsupervised.

The authors of the present contribution have been using Moodle-based
Multiple-Choice (MC) quizzes as a formative evaluation tool for many
academic years (2012-present) with the students enrolled in the
subject of General Physics of first year STEM (Science, Technology,
Engineering and Mathematics) degrees taught in the Faculty of Science
and Technology in the University of the Basque Country UPV/EHU.  At
the end of every topic of the subject, students answer 10 questions,
randomly extracted for every student and with answers randomized
within each question, from quite a large question-bank (around 60-70
per chapter) prepared by the lecturers.  The goal of these quizzes is
twofold: i) allow the students to self-evaluate their performance
along the semester; ii) give real-time feedback to the lecturer on the
performance of each student. A full description of this methodology
can be found in \cite{JOST-2016}.  A study of the results obtained
with this strategy demonstrates that those students regularly taking
and passing the tests increase their probability of passing the
course and, on average, obtain better final marks \cite{JOST-2016}.
Since 2016-2017 we have made them optional as a training tool
for a honest self-evaluation. In this case, even if the students are
not forced to solve the quizzes, they are aware that these are for
their own benefit, and the same conclusion achieved in the previous
study is drawn. Those who pass the tests, increase the probability
of passing the subject and obtain, on average, better marks
\cite{ICERI-2017}. We have also used the system at the MSc level
during two years.

If properly set out, the MC quizzes and Numerical questions in Moodle
can be a very valuable online tool for teaching and for (formative)
evaluation, as already shown in our previous studies. We list some of
the advantages of using them: they allow self-correction; questions
and answers can be randomized; they allow immediate or delayed
feedback; they can be made available only within a time-window; a
countdown for solving them can be set; questions can be classified in
categories and subcategories; questions can be exported from a course
to be imported back to another course in many different formats;
equations and images are allowed within the questions; and many
more. But probably one of most useful features of the Moodle question
bank is that when it is large enough, all students get randomly
assigned different questions of a similar level of complexity that
evaluate the same learning outcome. This guarantees that they can
not share the answers while they fill in the quiz. More importantly,
this ensures that all students are evaluated fully respecting the
principle of equal opportunities. Of course, this means that,
depending on the amount of students enrolled in the course, a huge
amount of questions must be generated. Thus, the workload upon
instructors increases dramatically.

The recent outbreak of the global COVID-19 pandemic in spring of 2020
has put the online teaching and learning on the focus of all
educational systems worldwide. Indeed, the closure of schools and
universities has literally obliged to adapt all the teaching-learning
process to the online world unexpectedly and in a very short
time. Lecturers and students have suddenly needed to modify the way
they teach and the way they learn, respectively. Moreover, due to the
long period of confinement ordered by most of the governments, a
face-to-face evaluation in most of the universities will not be
possible. Therefore, the final evaluation of the teaching-learning
process will also need to be done using online resources, with all the
drawbacks and difficulties this implies, such as students not being
under surveillance to avoid them sharing answers 
during the exam.

In this paper we introduce the \texttt{pygiftgenerator} module for
\texttt{python}~\cite{CS-R9526}. It is easy and intuitive
to use and it allows the instructors to prepare a large
amount of MC and numerical questions based on the set up of a single
problem (which can also include images). Since the system works by
allowing to change the values for the parameters
involved in the problem, many different instances of the same problem
can be easily produced in GIFT format. The output file in GIFT format
holds the questions and their corresponding answers, all of them
classified according to the Moodle categories specified by the
user. They can then be easily uploaded to the question-bank in the
Moodle platform of any course. If the instructors design carefully the
\emph{taxonomy} of categories to which the questions belong, the
automatic random selection of questions for every student from the
same or similar families of problems is straightforward and allows the
instructor to cover the syllabus of the course.

The paper is organized as follows: Section \ref{sec:methodology}
describes the methodology used by the module to generate
easy-to-import GIFT files in Moodle and Section
\ref{sec:implementation} outlines its implementation and
installation. Some applications to selected problems in different
fields of Physics are shown in Section \ref{sec:sample-problems}. We
finish with the discussion in Section \ref{sec:discussion} and
conclusions in Section \ref{sec:conclusions}.

\section{Methodology}
\label{sec:methodology}
The GIFT format~\cite{giftdef} is one of the formats used by the Moodle learning
environment to allow the preparation of questions for their batch
import into a given course. It is relatively simple for
the case of simple text questions, but it can also hold HTML tags, so
that basic text formatting and even remote url-defined files (images,
references) can be included in questions. \LaTeX\ code can also be
inserted in the GIFT files, so that complex equations are also
supported.

An important first step in this module is the preparation of variable
input parameters for every problem.  We show two different ways of
treating this.  In the first case, the parameter set is randomly
extracted from a combination of input lists specified by the user, which allows the
instructors to keep numbers used as input parameters \emph{nice
  looking} or inside a discrete set that is physically meaningful. The
way this system works by combining lists of parameters is illustrated
in file \exampleFile{EM\_DC\_circuit\_MC.py}
distributed with the module.


A second option is to produce the input parameters by means of calls
to random number generators from the program that is generating the
questions. Even in this case, as shown in
the example file 
\exampleFile{TH\_Blackbody\_Planck\_NQ.py}, some of the input
parameters may be preset through a list (representing a range of temperatures in this case) as in the previous example.


%

\begin{figure}
  \begin{center}
    \includegraphics[width=.9\linewidth]{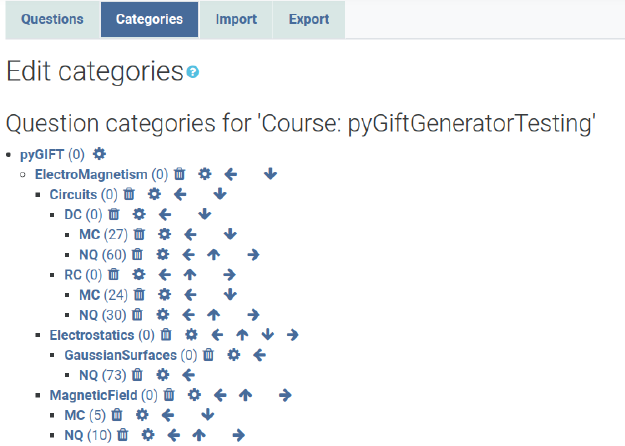}
    \caption{\label{fig:categories}Part of the category tree for the examples provided with the module
as imported in Moodle}
  \end{center}
\end{figure}

An important aspect previous to the design of the questions is the
preparation of a consistent tree of categories and subcategories which
allow the instructors to perform a proper classification of topics and
subtopics which aids in the random extraction of questions by the
Moodle quiz preparation system for each individual student.  These
categories are kept during the production of the GIFT code and are
imported by Moodle properly (see figure~\ref{fig:categories}). This is
important when random tests for students are designed on the basis of
diverse topics. They allow properly sampling the program covered
through the year and equally distributing different options to
every student.

There are many classes of questions available in Moodle, and all of them
can be imported by means of the GIFT format, such as MC, True-False,
Short Answer and Numerical to name a few. Considering that
we deal with exercises in the field of Physics, we have focused in
Numerical and MC types only, since we found them
to be the most interesting for carrying out exams during the
COVID-19 exam period. A similar structure of code can be
used for the other kind of questions, and it might even happen that we
support them in the future. However, in many cases (Short Answer,
Missing Words or Matching questions, for instance), the use of a
module such as \texttt{pygiftgenerator} wouldn't be effective at
all. The reason is that the use of \texttt{python} as a computer
language as we propose in this paper is perfect for questions in which
numerical computation is needed, since the answer can be recalculated
for different values of the parameters. This applies perfectly to
Numerical and MC questions. However, for Short Answer or Matching
questions, the use of a computer language doesn't draw a significant
advantage and using it would require much more work than the one
needed to write them directly in GIFT. Thus, the authors have
decided to exclusively support Numerical and MC questions.
In the latter case, the module ensures that questions with repeated answers are not included in the output GIFT file.

\begin{figure}
  \begin{center}
    \includegraphics[width=.9\linewidth]{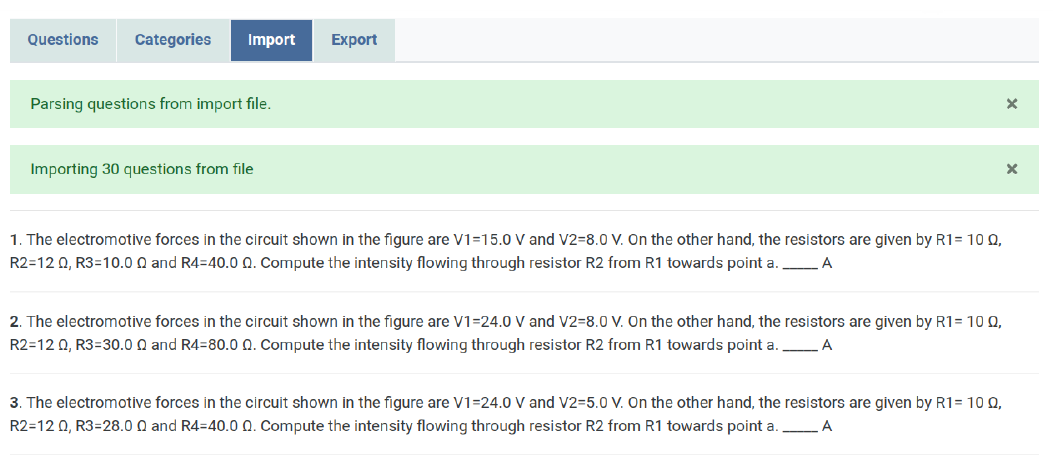}
    \caption{\label{fig:importing2}Some of the questions produced by
      the script in \exampleFile{EM\_DC\_circuit\_MC.py}.}
  \end{center}
\end{figure}

The GIFT file produced by a
\texttt{python} interpreter running a program built around the
\texttt{pygiftgenerator.py} module can be easily imported to the
question bank of a Moodle installation. 
A screenshot of the importing result is shown in figure~\ref{fig:importing2}.
Additionally, the questions are properly classified in the tree
categories and
subcategories initially designed by the instructors.


\section{Implementation}
\label{sec:implementation}
The module \texttt{pygiftgenerator} can be installed by means of a
standard \texttt{setup.py} installation script which uses the
\texttt{distutils} tools. 
The code has been tested with
versions 3.6, 3.7 and 3.8 of \texttt{python}.
It is open source distributed under the GPLv3 license and it can be obtained
at the \url{htps://gitlab.com/EHU/pygiftgenerator}
repository. 

\section{Results: Application to selected problems}
\label{sec:sample-problems}

We list here a selection of applications. Many more and their
corresponding example files can be found in the documentation
available with the module.

\subsection{Mechanics}
\label{sec:sample-mechanics}

\begin{figure} 
 \begin{center}
    \includegraphics[width=4cm]{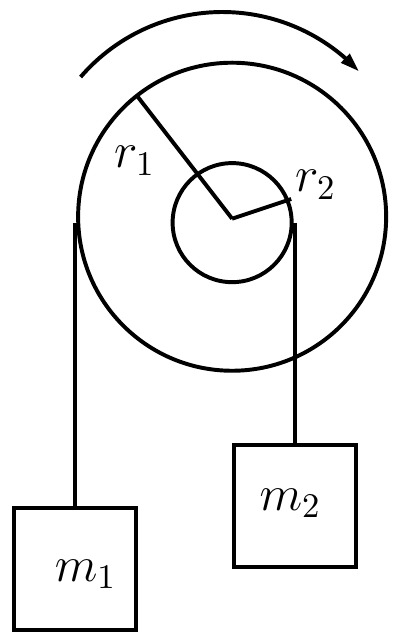}
    \caption{\label{fig:pulleys} Two coupled pulleys rotating about a
      common axis. A point mass hangs from a string wound to each of
      the pulleys. The arrow shows the positive direction of motion.}
 \end{center}
\end{figure}

The following example deals with the application of Newton's second
law to a system of pulleys and point masses. This is a classical
example covered in a General Physics course in first year
undergraduate studies related with the dynamics of rigid
bodies, since the pulleys we will deal with have a non-negligible
mass. Figure~\ref{fig:pulleys}
shows two coupled pulleys rotating about a common axis. The large
(small) one has radius $r_1$ ($r_2$) and a mass $m_1$ ($m_2$) hangs
from its rim with a mass-less string. When the system is released, the
ratio of parameters for each pulley  defines the direction of
rotation of the system.

The positive direction of rotation used in the problem
is marked with the arrow. This means that the accelerations of masses
$m_i$ and angular acceleration of the pulley-system will be positive
when the system rotates clockwise, and negative when the system
rotates anticlockwise.  With this criterion we can apply Newton's
second law to the masses attached to the strings $\left( \Sigma\vec{F}
\right)_i = m_i \vec{a}_i $ and calculate the torque on the pulleys
$\Sigma \vec{M} = I \vec{\alpha}$, with $a_i=\alpha r_i$ due to the
rotation-translation constraints of the strings around the pulleys.
%
Therefore the problem reduces to solving the following system of equations 
\begin{equation}
\left.
\begin{array}{lcr}
T_1 - m_1 g  &=& m_1 a_1 \\
a_1 &=& \alpha r_1
\end{array}
\right\}
\end{equation}
\begin{equation}
\left. 
\begin{array}{lcr}
m_2 g - T_2 &=& m_2 a_2 \\
a_2 &=& \alpha r_2 
\end{array}
\right\}
\end{equation}
\begin{equation}
\begin{array}{lcr}
 T_2 r_2 - T_1 r_1  &=& I \alpha 
\end{array}
\end{equation}

The examples in \texttt{pygiftgenerator} ask the students to solve the
problem for the following quantities:
\begin{enumerate}
\item The angular acceleration of the system, given by 
\begin{equation}  \label{Eq2:alpha}
\alpha=\frac{m_2r_2-m_1r_1}{m_1r_1^2+m_2r_2^2+I}g.
\end{equation}
\item The tension of the string attached to $m_1$, given by 
\begin{equation}
T_1 = m_1 (g+\alpha r_1),  
\end{equation}
with $\alpha$ from \Eref{Eq2:alpha}.
\item The tension of the string attached to $m_2$, given by 
\begin{equation}
T_2 = m_2 (g-\alpha r_2),  
\end{equation}
with $\alpha$ from \Eref{Eq2:alpha}.
\item The acceleration of any of the masses, given by
\begin{equation}
a_i = \alpha r_i,
\end{equation}
with $\alpha$ from \Eref{Eq2:alpha}.
\end{enumerate}

This is a clear example in which based on a single figure and slightly
changing the set up of the problem, one can prepare a large amount of
different questions involving the different physical quantities
represented by the equations above.  See example files
\exampleFile{ME\_coupled\_pulleys\_NQ.py} for
Numerical and \exampleFile{ME\_coupled\_pulleys\_MC.py} for
Multiple Choice examples.

%


\subsection{Electromagnetism}
\label{sec:sample-electromagnetism}
\subsubsection{Direct current circuit with multiple batteries and resistors}
\label{sec:dccircuit}
\begin{figure}
  \begin{center}
    \includegraphics[width=4cm]{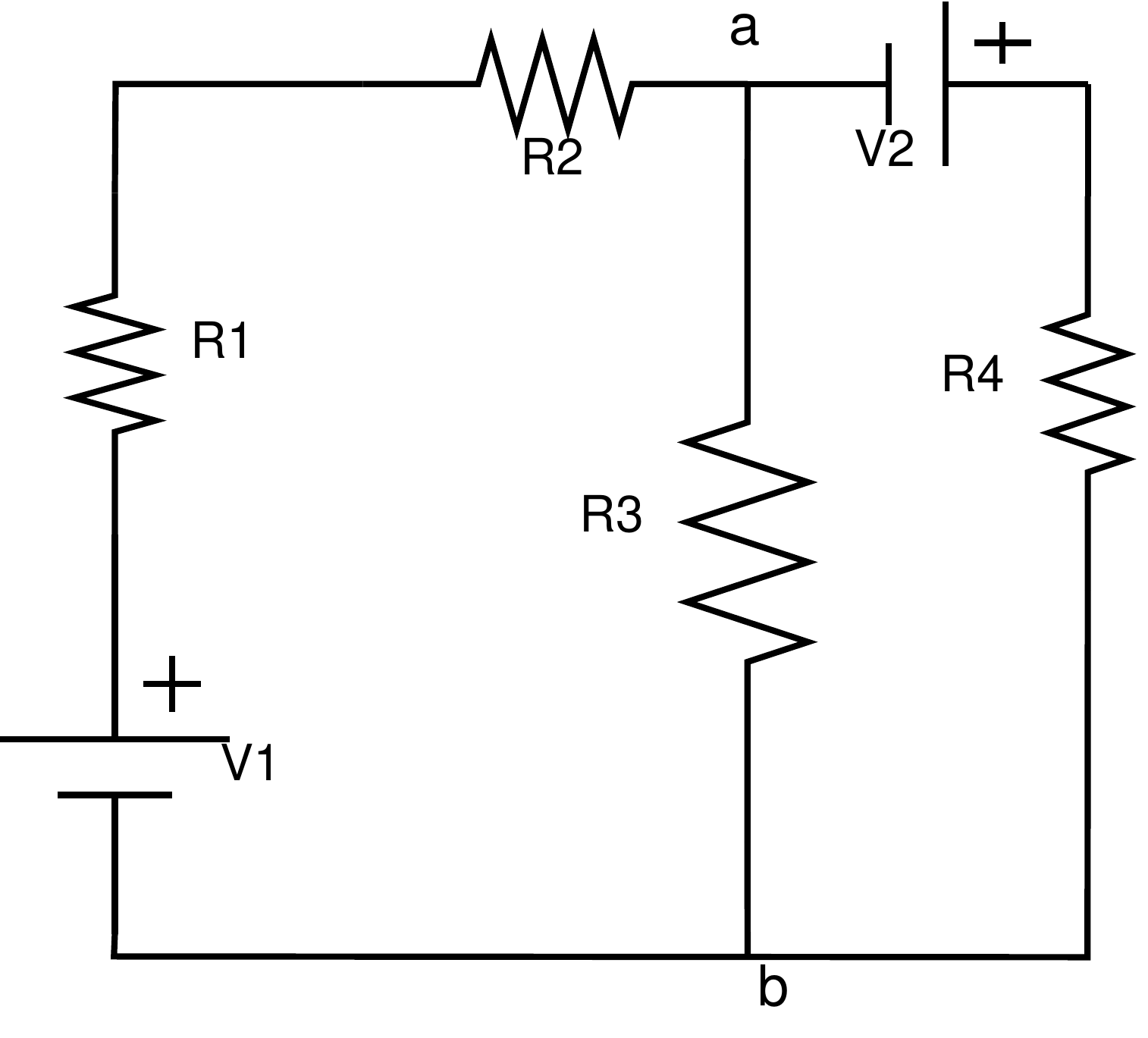}
    \caption{\label{fig:circuit1}Schematic representation of the DC
      circuit used in file \exampleFile{EM\_DC\_circuit\_MC.py}.}
  \end{center}
\end{figure}

This example covers one of the typical cases that can be found in
standard textbooks for first year undergraduate Physics in Physics or
Engineering degrees~\cite{tipler}. Figure~\ref{fig:circuit1} shows an
example where four resistors and two batteries are arranged forming a
network with two junctions and three possible loops. Two of the
resistors are fixed ($R_1=10\;\Omega$ and $R_2=12\;\Omega$) and the
input parameters that are varied in order to produce the different
questions are $R_3$, $R_4$ and the batteries' electromotive forces
$V_1$ and $V_2$.

The questions that are presented to the students range from finding
the intensities $I_2$ and $I_3$ flowing through resistors $R_2$ and
$R_3$ to calculating the power dissipated in resistor $R_3$. In all
cases, the circuit is solved by using the expressions:
\begin{equation}
  I_2=\frac{V_1\left(R_1+R_2\right)-
    \left(V_1+V_2\right)\left(R_1+R_2+R_3\right)}{\Delta}
\end{equation}
and
\begin{equation}
  I_3=\frac{\left(R_1+R_2\right)\left(V_1+V_2\right)-V_1\left(R_1+R_2+R_4\right)}{\Delta},
\end{equation}
with 
\begin{equation}
  \Delta=\left(\left(R_1+R_2\right)^2-\left(R_1+R_2+R_3\right)\left(R_1+R_2+R_4\right)\right).
\end{equation}

Solvers which produce questions regarding the intensities
$I_3$ and the power $P_3$ dissipated by $R_3$ are given in files
\exampleFile{EM\_DC\_circuit\_MC.py} 
and \exampleFile{EM\_DC\_circuit\_NQ.py} 
for MC and Numerical questions, respectively.



\subsubsection{Magnetic field produced by two parallel infinite wires}
  \begin{figure}
\begin{center}
    \includegraphics[width=5cm]{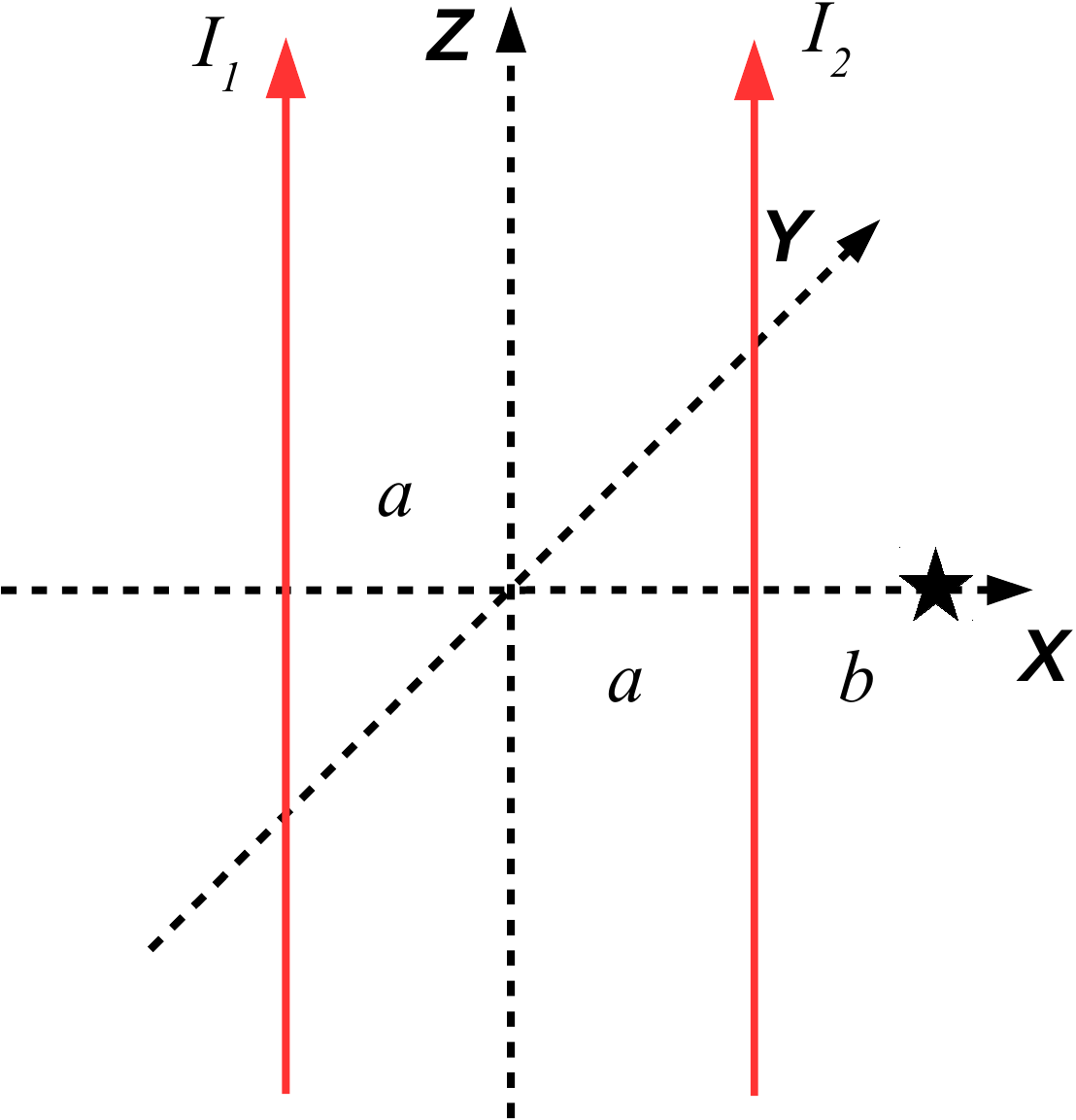}
    \caption{\label{fig:Bf}Parallel current-carrying wires producing
      the magnetic fields.}
\end{center}
    \end{figure}

 We first use the expression corresponding to the magnetic field
 produced by a wire ~\cite{giancoli,purcell_morin} (also known in
 electromagnetism as magnetic flux density~\cite{bleany}).  By using
 the superposition principle, we combine that expression with the
 field produced by a second parallel wire. Changing the variables defining the geometry of the problem
 and the intensities flowing through the conductors, many questions
 can be defined.  Currents as defined in figure~\ref{fig:Bf} are taken
 as positive and the student is required to calculate the total
 magnetic field at the point marked by a star, which is placed at a
 distance $b$ from the origin of coordinates
 and which is given by:
\begin{equation}
\vec{B}=\frac{\mu_0}{2\pi}\left(\frac{I_1 s_1}{b+a}+\frac{I_2s_2}{b-a}\right)\hat\jmath
\end{equation}
with $s_1$ and $s_2$ coefficients valued 1 or -1 according to the
orientation of currents $I_1$ and $I_2$ and $\mu_0$ the permeability
of free space. This example is coded in file
\exampleFile{EM\_Magnetic\_Field\_NQ.py} as a Numerical question
and in file \exampleFile{EM\_Magnetic\_Field\_MC.py}
as Multiple Choice.

%

\subsection{Thermodynamics}
\subsubsection{Equation of state of air}
\label{sec:es}
Following relevant literature~\cite{petty}, we can build questions related to the
determination of the density of dry air 
$\rho_{\rm{d}}={P}/({R_{\rm{d}}T})$, 
with
$R_{\rm{d}}=287$ J kg\unitto{-1} K\unitto{-1} the gas constant corresponding
to the mixture of dry air, $P$ pressure in Pa and $T$ air temperature
in K.

The density of moist air $\rho_{\rm{m}}=(P-e)/(R_{\rm{d}}T)+{e}/({R_{\rm{v}}T})$  
is
computed from the mixture of dry air and water vapour, with $R_{\rm{v}}=461$
J kg\unitto{-1} K\unitto{-1} the gas constant corresponding to water
vapour and $e$ the partial pressure corresponding to water vapour in
the moist air measured in Pa.

The virtual temperature of air $T_{\rm{v}}$ represents the temperature at
which dry air has the same density as a given parcel of moist air and
is computed by means of the expression:
\begin{equation}
T_{\rm{v}}=\frac{T}{1-\frac{e}{P}\left(1-\frac{R_{\rm{d}}}{R_{\rm{v}}}\right)}.
\end{equation}
The expressions above can be combined with \Eref{ec:es} to
calculate the saturation pressure of water vapour $e_{\rm{s}}$ in Pa as a
function of temperature through a common approximation~\cite{wallace}:
\begin{equation}
e_{\rm{s}}=611\exp\left[ \frac{L}{R_{\rm{v}}}\left( \frac{1}{273}-\frac{1}{T}\right)\right] 
\label{ec:es}
\end{equation}
with $L=2.5\times 10^6$ J kg\unitto{-1} the latent heat of evaporation
of water. This allows to compute relative humidity defined from water
vapour mixing ratio 
$w={\rho_{\rm{v}}}/{\rho_{\rm{d}}}$ as:
\begin{equation}
r=100\frac{w}{w_{\rm{s}}}=100\frac{e}{e_{\rm{s}}}\left(\frac{P-e_{\rm{s}}}{P-e}\right).
\end{equation}


\subsubsection{Adiabatic evolution of unsaturated air}
In the last example in this section, potential temperature
\begin{equation}
\label{eq:theta}
\theta=T\left( \frac{P_0}{P}\right)^{R_{\rm{d}}/c_{\rm{p}}}
\end {equation}
is computed to determine the adiabatic evolution of air, with $c_{\rm{p}}$ the
specific heat of air at constant pressure. These questions must use
pressures and temperatures at given ranges of temperature and pressure
at the atmosphere to be meaningful, as shown in example
\exampleFile{TH\_adiabatic\_process\_NQ.py}.


\subsection{Modern Physics}
\subsubsection{Relativistic collisions of particles with photons}
In this example, it is considered that a photon with energy $E$
collides against a particle at rest with mass $m=\frac{E_0}{c^2}$ and
we want to obtain the energy $E^*_{\rm{m}}$ and linear momentum of the
particle as seen from the center of mass reference frame (CM-RF). The
velocity of the CM-RF is given by:
\begin{equation}
  v=\frac{E c}{E+E_0},
\end{equation}
which yields a value of
\begin{equation}
  \gamma=\frac{E+E_0}{\sqrt{E_0^2+2E E_0}}
\end{equation}
for the CM-RF Lorentz's factor. Thus, the particle's
energy after the collision, as seen from the CM-RF is
\begin{equation}
  E^*_{\rm{m}}=\left( E + E_0\right)\sqrt{\frac{E_0}{E_0+2E}}
\end{equation}
and its linear momentum
\begin{equation}
  p^*_{\rm{m}}=-\frac{E}{c}\sqrt{\frac{E_0}{E_0+2E}}.
\end{equation}
Example file
\exampleFile{MP\_SRelativity\_CMCollision\_NQ.py} shows the
code that produces results for collisions between photons and particles with different energies. As an improvement to previous examples, in
this case, the output is written in an ASCII file by the code, and no
redirection is needed. In order to get that done, the additional
parameter \verb+ofile+ is added to the constructor of the question
objects.


\subsubsection{Photoelectric effect}
Let the photoelectric work function of a metal be $W$. If light having
a wavelength $\lambda$ shines on the surface, photo-electrons with kinetic energy $K$ will be
emitted from the surface  provided that the
energy of the incident beam is larger than the work function of the
metal. These three magnitudes are related by
\begin{equation}
\frac{hc}{\lambda} = K + W,
\end{equation}
where $c$ is the speed of light and $h$ is Planck's constant.  The
photoelectric threshold wavelength for the surface is the minimum
wavelength needed to eject electrons from the surface:
\begin{equation}
\lambda_0= \frac{hc}{W}.
\end{equation}

The maximum velocity of an ejected electron is:
\begin{equation}
v_{\rm{max}}=\sqrt{\frac{2 K_{\rm{max}}}{m_{\rm{e}}}}=\sqrt{\frac{2(hc/\lambda-W)}{m_{\rm{e}}}} \,,
\end{equation}
where $m_{\rm{e}}$ is the mass of the electron
and its corresponding de Broglie wavelength reads
\begin{equation}
\lambda=\frac{h}{p}=\frac{h}{m_{\rm{e}} v_{\rm{max}}}.
\end{equation}

The example in file
\exampleFile{MP\_photEffect\_workfunc\_NQ.py} allows
to generate questions for the excitation threshold wavelength while the
code provided in file
\exampleFile{MP\_photEffect\_ejectedElectrons\_NQ.py} can be used to
either obtain the maximum velocity of the ejected electrons or their
deBroglie wavelength.

\section{Discussion}
\label{sec:discussion}
There are some question types such as Simple Calculated, Calculated
and Calculated MC questions supported in Moodle. Thus, it can be fair
to question the need of a module like the one presented in this
paper. As we see it, performing the computations in an external
language such as \texttt{python}, allows to achieve at least four
objectives. First, the degree of complexity of the problems which can
be handled can be extended beyond the ones available in the inner core
of Moodle.

Second, as shown in some of the examples above, particularly the ones
involving the case of the coupled pulleys
(Section~\ref{sec:sample-mechanics})
or the DC circuit (Subsection~\ref{sec:dccircuit}), different versions
of similar problems can be prepared using a single set of
parameters. With little effort, the amount of generated different
problems increases, even if most of them can be solved by applying the same
concepts.  If the amount of time given for solving the online
questionnaire is properly calibrated, the results obtained from the test should be
trustworthy as reflecting students' knowledge even if the quiz is
unsupervised.

Third, the fact that many parts of the solution are common for the
same set of parameters and different questions (see in particular the
cases of the coupled pulleys or DC circuit) makes it easier to produce
the set of questions by means of the solution proposed in this paper,
since these variables are shared during the generation of the
questions in the same piece of code, avoiding thus potential coding
errors that might happen if the questions were prepared independently.

Fourth, \texttt{python} is a full featured computer language with
access to many external modules and packages and with the ability to build
complex programs. The capability to write \LaTeX\ code in GIFT questions
allows the instructors to build complex equations (determinants or
systems of equations for instance). This, combined with the linear
algebra routines provided by \texttt{numpy} allow to use this system
to solve problems of linear algebra of higher dimension which would be
very complex to code directly in Moodle. Additionally, the use of
\texttt{python} as the computing core allows the
instructors to use external auxiliary functions (see the example
\exampleFile{TH\_relative\_humidity\_NQ.py} for calculating
the saturation pressure of water vapour) that are used once and again
during the computation of many quantities involved in the solutions of
the problems at hand. Having to rewrite them in every question would
be very error-prone and time-consuming.

There are many learning environments, such as Moodle, Blackboard or
Canvas, to name a few, which support the use of calculated
questions. We are addressing in this paper the platform that our
university provides to us, but after a revision of the features of
calculated (\emph{formula}) questions in these platforms, the previous
statements still stand true. The solution provided here allows more
complex computations with better reliability, so that the same
methodology could probably be extended to formats allowed by those
platforms for importing questions.

We have provided interfaces for producing Numerical questions and
Numerical MC questions, which are the ones that we find more relevant
from the point of view of teaching Physics and the massive production
of questions for a question bank which should allow individual and
independent tests for every student. There are, of course, other
options in Moodle such as MC text questions which allow challenging
conceptual tests in the field of Physics. However, from the point of
view of producing those questions, our experience from the past years
shows that MC conceptual text questions are better produced directly
on the Moodle interface and that a \texttt{python} program would only
complicate the steps needed to arrive to the final product (a GIFT
file to be imported to the question bank). An additional advantage of
using the GIFT format for importing the results into Moodle is that it
is a simple ASCII-based format which can be easily read by humans
before uploading the file to Moodle, so that inconsistencies can be
easily detected.  In this first release of the software, we have
selected the GIFT format for allowing an easy upload of the questions
to the Moodle server.  Depending on the configuration of the server,
this may make the inclusion of figures rather cumbersome.
Images can also be encoded using \texttt{base64} package into the GIFT file,
as shown in \exampleFile{EM\_circuit2\_NQ.py}. However, this leads to very large and difficult to read GIFT files.
Probably,
future releases of \texttt{pygiftgenerator} will also export questions
in the XML Moodle format to ease the upload of figures and other
resources.

We have tested this solution for preparing an online exam for 220 students during the first weeks of COVID-19 lockdown and results seem promising, but a full analysis of the results is not possible, since we lack a control group.

\section{Conclusions}
\label{sec:conclusions}
In this contribution we present a \texttt{python}-based flexible
module for producing a large amount of Numerical and MC Numerical
questions oriented to evaluating students online. This is very
relevant currently, during the lockdown due to the COVID-19 pandemic,
which has enforced the closure of most schools and universities.  In
particular, new online tools need to be designed to evaluate students'
learning process. If these tools are meant to substitute on-site
supervised written exams, they require a minimum guarantee that
students will not be able to share answers while solving the exam
because of the fact that no one is supervising them in a room.  This
\texttt{python} module allows to produce a large question bank. All the
questions are different but can be classified into categories and
types.  This way, the probability of students getting quizzes with
exactly the same questions is very low, and can be made negligible
just by increasing the amount of items prepared. The module is
quite flexible, as shown in this contribution and can be used to
cover different branches of Physics, with examples selected in this
contribution covering many different topics, such as Dynamics,
Thermodynamics, Electromagnetism, Special Relativity and Quantum
Mechanics.
We have selected \texttt{python} because it is a simple programming
language. It is very friendly for people who are not programming
experts. Its learning curve is not too steep. Besides that, the module
presented here is quite easy to use, which makes us think that it can
be used by instructors worldwide without a significant inversion of
time for learning its working.
However, the system is powerful; the examples provided in the
documentation of the module show that it can produce many different
questions from the same set of parameters with a unique
identification and organized in categories. Equations
written in \LaTeX\ or even figures can be imported by Moodle from the
GIFT files generated.

\section*{References}
\bibliographystyle{unsrt}
\bibliography{pygift}

\begin{thebibliography}{10}

\bibitem{ECTS-2017}
Directorate-General for Education {Youth, Sport, and} Culture
  (European~Commission).
\newblock {ECTS User's} guide 2015, 2017 (accesed April 8, 2020).

\bibitem{BP-2015}
European Commission/EACEA/Eurydice.
\newblock The {E}uropean {H}igher {E}ducation {A}rea in 2015: {B}ologna
  {P}rocess {I}mplementation {R}eport.
\newblock Luxembourg: Publications Office of the European Union.

\bibitem{usreport}
B.~Means, Y.~Toyama, R.~Murphy, M.~Bakia, and K.~Jones.
\newblock Evaluation of evidence-based practices in online learning. a
  meta-analysis and review of online learning studies.
\newblock Technical report, US Department of Education, 2010.

\bibitem{Kozma-1994}
R.B. Kozma.
\newblock Will media influence learning: Reframing the debate.
\newblock {\em Educational Technology Research and Development}, 42(2):7--19,
  1994.

\bibitem{Clark-1994}
R.~E. Clark.
\newblock Media will never influence learning.
\newblock {\em Educational Technology Research and Development}, 42(2):21--29,
  1994.

\bibitem{Borondo-2014}
J.~Borondo, R.M. Benito, and J.C. Losada.
\newblock Adapting physics courses in an engineering school to the b-learning
  philosophy.
\newblock {\em European Journal of Engineering Education}, 39(5):496--506,
  2014.

\bibitem{Freeman-2014}
S.~Freeman, S.L. Eddy, M.~McDonough, N.~Okoroafor M.K.~Smith, H.~Jordt, and
  M.P. Wenderotha.
\newblock Active learning increases student performance in science,
  engineering, and mathematics.
\newblock {\em Proceedings of the National Academy of Sciences of the United
  States of America}, 111(23):8410--8415, 2014.

\bibitem{ICERI-2017}
J.~S\'aenz, G.A. L\'opez, J.~Martinez-Perdiguero, I.~Alonso, A.~Leonardo, and
  I.G. Gurtubay.
\newblock The use of {M}oodle quizzes by students of first year {P}hysics in
  the university: mandatory versus optional quizzes.
\newblock In {\em ICERI2017 Proceedings}, 10th International Conference of
  Education, Research and Innovation, pages 5469--5472. IATED, 16th-18th
  November 2017.

\bibitem{IATED-2018}
I.G. Gurtubay, J.~S{\'a}enz, G.A. L{\'o}pez, S.J. Gonz{\'a}lez-Roj{\'i},
  I.~Unzueta, P.~Garcia-Goiricelaya, I.~Alonso, A.~Leonardo, and
  J.~Martinez-Perdiguero.
\newblock Learning {P}hysics from wrong preconceptions through daily-life
  related experiments.
\newblock In {\em INTED2018 Proceedings}, 12th International Technology,
  Education and Development Conference, pages 5703--5711. IATED, 5th-7th March
  2018.

\bibitem{EDULEARN-2018}
I.G. Gurtubay, J.~S{\'a}enz, I.~Alonso, J.~Lafuente-Bartolome, G.A. L{\'o}pez,
  J.~Martinez-Perdiguero, and A.~Leonardo.
\newblock Training {P}hysics {student's} intuition through daily-life related
  experiments.
\newblock In {\em EDULEARN18 Proceedings}, 10th International Conference on
  Education and New Learning Technologies, pages 5742--5750. IATED, 2nd-4th
  July 2018.

\bibitem{EuroSTL-2019}
I.~G. Gurtubay, J.~S{\'a}enz, and G.A. L{\'o}pez.
\newblock Laurel vs hardy: a {PBL} activity for motivating first year
  university students in {P}hysics.
\newblock In {\em Third {EuroSoTL Conference}: {E}xploring new fields through
  the scholarship of teaching and learning}, pages 206--213. Servicio Editorial
  de la Universidad del Pa\'{\i}s Vasco, 2019.

\bibitem{moodle}
Moodle {P}roject, 2020, (accessed April 7, 2020).

\bibitem{broadbent_2018}
J.~Broadbent and M.~{F}uller{-T}yszkiewicz.
\newblock Profiles in self-regulated learning and their correlates for online
  and blended learning students.
\newblock {\em Education Tech Research Dev}, 66:1435–1455, 2018.

\bibitem{liagkou_2018}
V.~Liagkou and C.~Stylios.
\newblock A trustworthy and privacy preserving model for online competence
  evaluation system.
\newblock In {\em Contemporary Complex Systems and Their Dependability},
  Proceedings of the Thirteenth International Conference on Dependability and
  Complex Systems DepCoS-RELCOMEX, July 2-6, 2018, Brun{\'o}w, Poland, page
  338–347. Springer, 2018.

\bibitem{corrigan_et_al_2015}
H.~Corrigan{-G}ibbs, N.~Gupta, C.~Northcutt, E.~Cutrell, and W.~Thies.
\newblock Deterring cheating in online environments.
\newblock {\em ACM Trans. Comput.-Hum. Interact.}, 22, 2015.

\bibitem{sithara_et_al_2019}
S.~H. P.~W. Gamage, J.~R. Ayres, M.~B. Behrend, and E.~J. Smith.
\newblock Optimising {M}oodle quizzes for online assessments.
\newblock {\em International Journal of {STEM} Education}, 6:27, 2019.

\bibitem{JOST-2016}
G.A. L\'opez, J.~S\'aenz, A.~Leonardo, and I.G. Gurtubay.
\newblock Use of the {M}oodle platform to promote an ongoing learning when
  lecturing {G}eneral {P}hysics in the {P}hysics, {M}athematics and
  {E}lectronic {E}ngineering programmes at the {U}niversity of the {B}asque
  {C}ountry ({UPV/EHU}).
\newblock {\em Journal of Science Education and Technology}, 25(4):575--589,
  2016.

\bibitem{CS-R9526}
G.~van Rossum.
\newblock Python tutorial.
\newblock Technical Report CS-R9526, Centrum voor Wiskunde en Informatica
  (CWI), Amsterdam, May 1995.

\bibitem{giftdef}
Moodle Project.
\newblock {\em Moodle Docs 3.8. GIFT format}, 2018 (accessed April 7, 2020).

\bibitem{tipler}
P.~A. Tipler and G.~Mosca.
\newblock {\em Physics for Scientists and Engineers}.
\newblock W. H. Freeman and Co., New York, NJ, USA, 6th edition, 2004.

\bibitem{giancoli}
D.~C. Giancoli.
\newblock {\em Physics for Scientists \& Engineers with Modern Physics}.
\newblock Pearson Education Limited, Harlow, UK, 2014.

\bibitem{purcell_morin}
E.~M. Purcell and D.~J. Morin.
\newblock {\em Electricity and Magnetism}.
\newblock Cambridge University Press, Cambridge, UK, 3rd edition, 2013.

\bibitem{bleany}
B.~I. Bleaney and B.~Bleaney.
\newblock {\em Electricity and Magnetism. Volume 1}.
\newblock Oxford University Press, Oxford, UK, 3rd edition, 2013.

\bibitem{petty}
G.~W. Petty.
\newblock {\em A First Course in Atmospheric Thermodynamics}.
\newblock Sundog Publishing, Madison, WI, USA, 2008.

\bibitem{wallace}
J.~M. Wallace and P.~V. Hobbs.
\newblock {\em Atmospheric Science. An Introductory Survey}.
\newblock Elsevier, Burlington, MA, USA, 2nd edition, 2006.

\end{thebibliography}
\end{document}